\input{aipcheck}

\documentclass[
    ,final            
  ]
  {aipproc}
\usepackage{graphicx}
\usepackage{amsfonts}
\usepackage{amssymb}
\usepackage{amsmath}
\usepackage{mathptmx}
\usepackage{rsfs}
\usepackage{bm}
\usepackage{ulem}
\usepackage{color}
\usepackage{shortvrb}

\layoutstyle{8x11single}

\begin{document}

\title{Reformulations of the Yang-Mills theory 
\\
toward quark confinement and mass gap}

\author{Kei-Ichi Kondo}{
  address={Department of Physics, Graduate School of Science, Chiba University, Chiba 263-8522, Japan}
}

\author{Seikou Kato}{
  address={Fukui National College of Technology, Sabae 916-8507, Japan}
}

\author{Akihiro Shibata}{
  address={Computing Research Center, High Energy Accelerator Research Organization (KEK), Tsukuba  305-0801, Japan}
}

\author{Toru Shinohara}{
  address={Department of Physics, Graduate School of Science, Chiba University, Chiba 263-8522, Japan}
}

\begin{abstract}
We propose the reformulations of the $SU(N)$ Yang-Mills theory toward quark confinement and mass gap. In fact, we have given a new framework for reformulating the $SU(N)$ Yang-Mills theory using new field variables.  This includes the preceding works given by Cho, Faddeev and Niemi, as a special case called the maximal option in our reformulations. The advantage of our reformulations is that the original non-Abelian gauge field variables can be changed into the new field variables such that one of them called the restricted field gives the dominant contribution to quark confinement in the gauge-independent way.  Our  reformulations can be combined with the $SU(N)$ extension of the Diakonov-Petrov version of the non-Abelian Stokes theorem for the Wilson loop operator to give a gauge-invariant definition for the magnetic monopole in the $SU(N)$ Yang-Mills theory without the scalar field.  In the so-called minimal option, especially, the restricted field is non-Abelian  and involves the non-Abelian magnetic monopole with the stability group $U(N-1)$. This suggests the non-Abelian dual superconductivity picture for quark confinement. This should be compared with the maximal option: the restricted field is Abelian  and involves only the Abelian magnetic monopoles  with the stability group $U(1)^{N-1}$, just like the Abelian projection. We give some applications of this reformulation, e.g., the stability for the homogeneous chromomagnetic condensation of the Savvidy type, the large N treatment for deriving the dimensional transmutation and understanding the mass gap, and also the numerical simulations on a lattice which are given by Dr. Shibata in a subsequent talk.

\end{abstract}

\maketitle

\def\slash#1{\not\!#1}
\def\slashb#1{\not\!\!#1}
\def\slashbb#1{\not\!\!\!#1}


\section{Introduction}

\begin{figure}[tb]
\includegraphics[height=4.0cm]{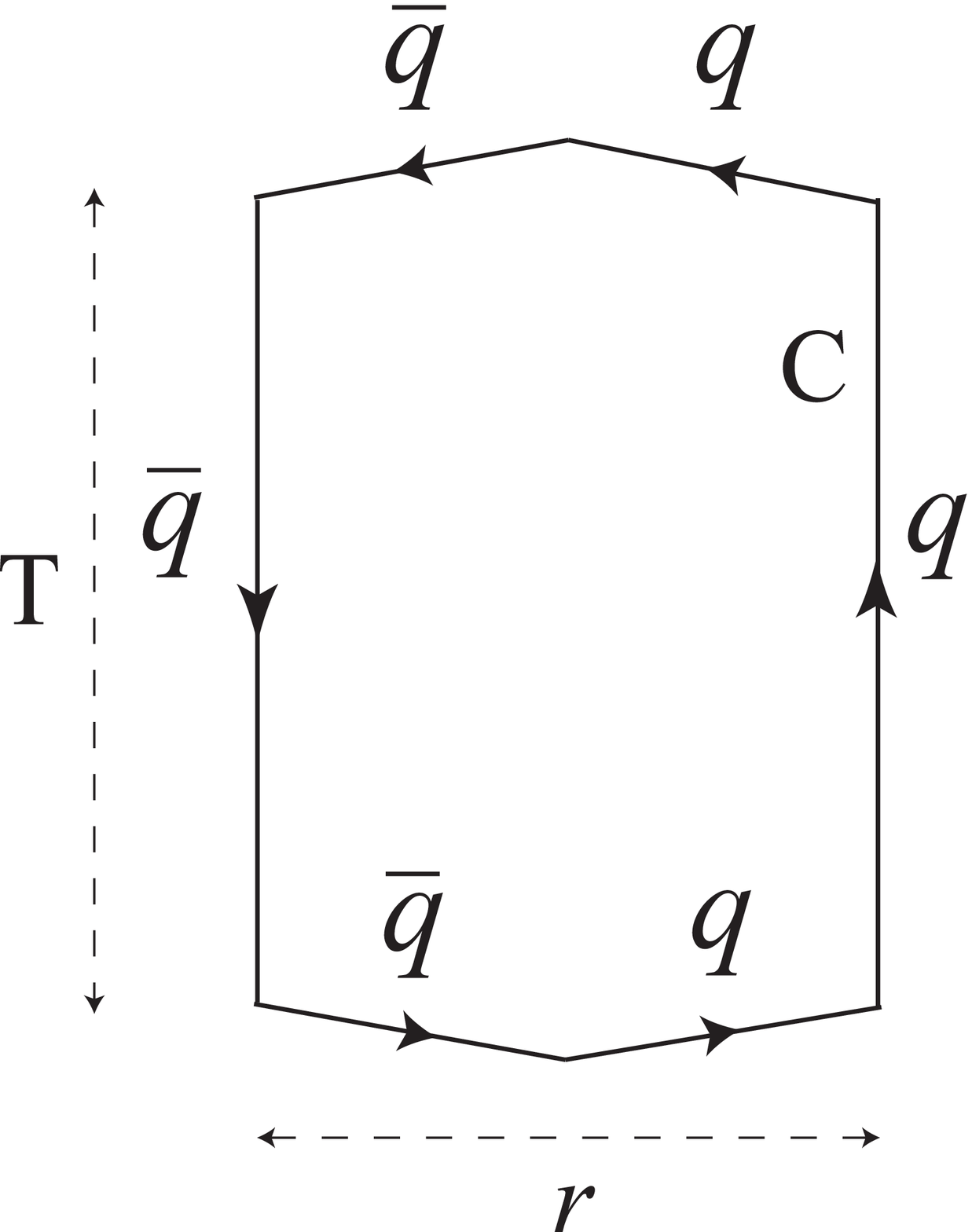}
\quad\quad\quad\quad
\includegraphics[height=4.0cm]{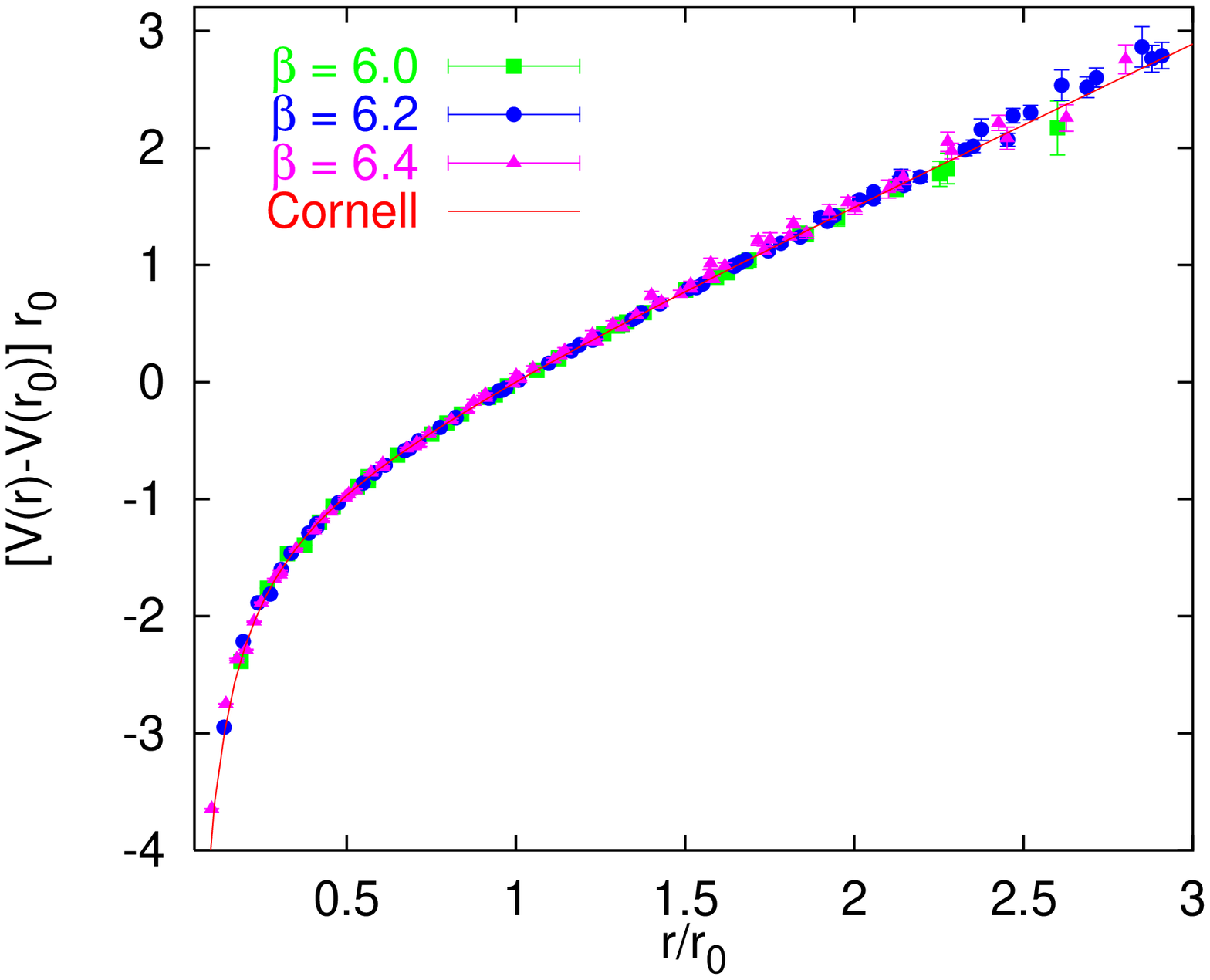}
\caption{
(Left) The Wilson loop for a rectangular loop $C$.
(Right) The static quark-antiquark potential $V(r)$ as a function of the  distance $r$ in SU(3) Yang-Mills theory, which is obtained by numerical simulations in the framework of lattice gauge theory.
Note that the potential is normalized so that $V(r_0)=0$ and $\beta=2N/g_{{}_{\rm YM}}^2$ for $SU(N)$.
See 
G.S. Bali,[hep-ph/0001312], Phys.Rept.{\bf 343}, 1 (2001).
}
\label{fig:potential}
\end{figure}

We follow the Wilson criterion for quark confinement, i.e., area law of the Wilson loop [Wilson (1974)]\cite{Wilson74}. 
Therefore, we start from the Wilson loop operator. 
For a given closed path $C$, we define the \textbf{Wilson loop operator} 
$W_C[\mathscr{A}]$ for the \textbf{non-Abelian Yang-Mills field} 
$\mathscr{A}_\mu(x)$ by [Yang $\&$ Mills (1954)]\cite{YM54}
\begin{align}
  W_C[\mathscr{A}]  
:=& {\rm tr} \left[ \mathscr{P} \exp \left\{ ig_{{}_{\rm YM}} \oint_{C} dx^\mu \mathscr{A}_\mu(x) \right\} \right]/{\rm tr}({\bf 1})  
 ,
\quad 
  \mathscr{A}_\mu(x)=\mathscr{A}_\mu^A(x) T_A  ,
\end{align}
where $\mathscr{P}$ denotes the \textbf{path-ordering} prescription. 
In the Yang-Mills theory, we consider the \textbf{Wilson loop average}  $W(C)$, i.e., a vacuum expectation value of the Wilson loop operator $W_C[\mathscr{A}]$ for a closed loop $C$:
	\begin{align}
		W(C) = \langle W_C[\mathscr{A}] \rangle_{\rm YM} .
	\end{align}
For a rectangular loop $C$ of side lengths $T$ and $r$, 
the  {Wilson loop average} $W(C)$ is related to a \textbf{static quark-antiquark potential} $V_{q\bar q}(r)$ as (See the left panel of Fig.~\ref{fig:potential})
	\begin{align}
		W(C) \sim \exp[-TV_{q\bar q}(r)] ,   \quad ( T \gg  r  ) .
	\label{W(C)} 
	\end{align}
$W_C[\mathscr{A}]$ is gauge invariant. 
Therefore, $V_{q\bar q}(r)$ is obtained in the gauge-independent way from 
	\begin{align}
		V_{q\bar q}(r) = \lim_{T\to \infty} \frac{-1}{T} \ln W(C) .
	\end{align}

\noindent
The numerical simulations exhibit that the static quark-antiquark potential
$V_{q\bar q}(r)$ is well fitted by the form of the Cornell type: 
Coulomb+{Linear} (See the right panel of Fig.~\ref{fig:potential})
\begin{equation}
 V_{q\bar q}(r) = - \frac{\alpha}{r} + {  \sigma r } + c ,
\nonumber
\end{equation}
wjth the three parameters of different dimensions, 
$\sigma$: string tension [mass$^2$], $\alpha$: dimensionless [mass$^0$], and $c$: [mass$^1$]. 
\\
$\bullet$ $\sigma \not= 0$ confinement $V_{q\bar q}(r) \rightarrow \infty$ as $r \rightarrow \infty$
\\
$\bullet$ $\sigma = 0$ deconfinement $V_{q\bar q}(r) < \infty$ as $r \rightarrow \infty$
\\
The emergence of the dimensionful string tension $\sigma$ is quite nontrivial, since the Yang-Mills theory includes the dimensionless parameters alone in the classical level.

\begin{figure}[ptb]
\includegraphics[scale=0.5]{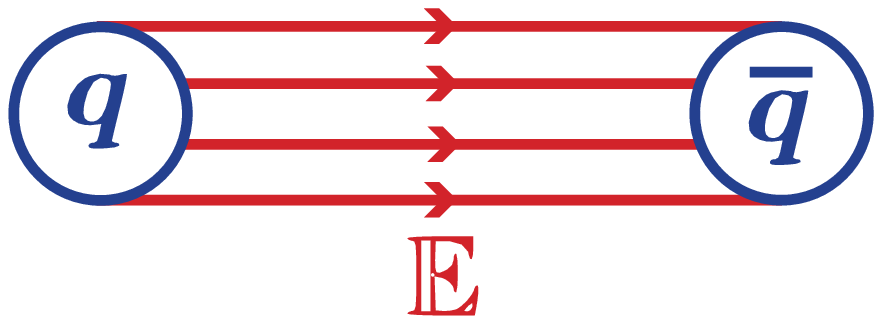}
\quad $\leftarrow$ dual $\rightarrow$ \quad 
\includegraphics[scale=0.5]{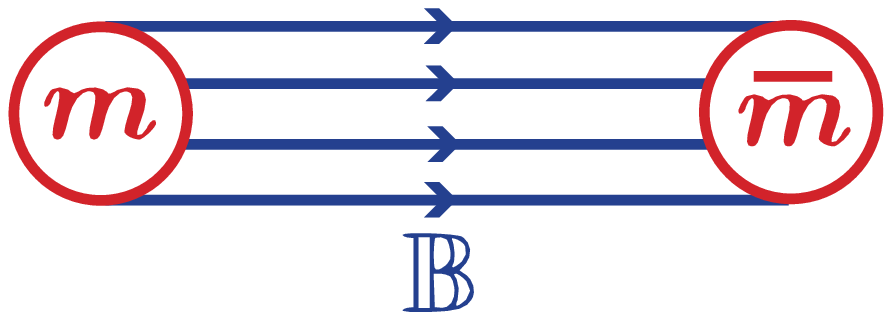}
\caption{
The electro-magnetic duality: electric charge is replaced by the magnetic charge, and the electric field is replaced by the magnetic field, and vice versa. 
}
\label{fig:duality}%
\end{figure}

A promising scenario for understanding quark confinement is  called the 
\textbf{dual superconductor hypothesis for quark confinement} based on the electro-magnetic duality (See Fig.~\ref{fig:duality}) proposed by  
[Nambu (1974), 't Hooft (1975), Mandelstam (1976), and Polyakov (1975,1977) \cite{dualsuper}.
The key ingredients for the dual superconductivity are as follows. 
See \cite{CP97,Greensite03} for reviews. 
\begin{itemize}
\item
\textbf{dual Meissner effect}
\\
In the dual superconductor,   chromoelectric flux   is squeezed into  tubes.
\\{ [$\leftarrow$  In the ordinary superconductor, magnetic flux is squeezed into tubes]} 

\item
\textbf{condensation of chromomagnetic monopoles  }
\\
The dual superconductivity is caused by condensation of chromomagnetic monopoles.
\\{
[$\leftarrow$ The ordinary superconductivity is cased by  condensation of electric charge into Cooper pairs. ]}

\end{itemize}

In order to establish the dual superconductivity, we must answer the following questions:

\noindent
* How to introduce {  magnetic monopoles in the Yang-Mills theory without scalar fields}?  
[This should be compared with the 't Hooft-Polyakov magnetic monopole.]
\\
\noindent
* How to define the {  duality in the non-Abelian gauge theory}?

\section{Non-Abelian Stokes theorem (1)}

\begin{figure}[tbp]
\includegraphics[height=3.0cm]{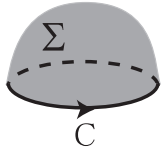}
\caption{
A closed loop $C$ for defining the Wilson loop operator and the surface $\Sigma_{C}$ whose boundary is given by the loop $C$. 
}
\label{fig:W_loop-Sigma}
\end{figure}

In order to answer the first question, we consider how the Wilson loop can be related to the magnetic monopole.

First, we consider the Abelian case. 
The Abelian Wilson loop operator $W_C[A]$ for a loop $C$ is cast into the surface integral over the surface $\Sigma_{C}$ bounded by $C$ using the \textbf{Stokes theorem}:(See Fig.~\ref{fig:W_loop-Sigma})
\begin{align}
  W_C[A]  
 =& \exp \left[  ie \oint_{C} dx^\mu A_\mu \right] 
\Longrightarrow 
W_C[A]  =   \exp \left[  ie \int_{\Sigma_{C}: \partial \Sigma_{C}=C} dS^{\mu\nu}(x(\sigma)) F_{\mu\nu}(x(\sigma))  \right] . 
\end{align}
\begin{picture}(0,0)
\end{picture}
Introduce the antisymmetric tensor $\Theta^{\mu\nu}_{\Sigma_{C}}$ called the \textbf{vorticity tensor} with the support only on the surface $\Sigma_{C}$ bounded by the loop $C$:
\begin{align}
   \Theta^{\mu\nu}_{\Sigma_{C}} (x) 
:=&   \int_{\Sigma_{C}: \partial \Sigma_{C}=C}  d^2 S^{\mu\nu}(x(\sigma)) \delta^D(x-x(\sigma)) .
\end{align}
\noindent
Then the surface integral is rewritten into the spacetime integral over the $D$-dimensional spacetime:
\begin{align}
W_C[A]  =&   \exp \left\{  ie (\Theta_{\Sigma_{C}},F)  \right\} , 
 \quad 
 (\Theta_{\Sigma_{C}},F) := \int d^Dx  \Theta^{\mu\nu}_{\Sigma_{C}} (x) F_{\mu\nu}(x) .
\end{align}
The Hodge decomposition can be used to define the electric current $j$ and the \textbf{magnetic current} $k$:
\begin{align}
  W_C[A]  
 =& \exp \left\{  ie  (N_{\Sigma_{C}},j) + ie (\Xi_{\Sigma_{C}}, k)   ] \right\} , 
 \quad
 N_{\Sigma_{C}} :=  \delta  \Delta^{-1}  \Theta_{\Sigma_{C}} ,
 \quad
 \Xi_{\Sigma_{C}} :=  
  \delta  \Delta^{-1}  {}^{\displaystyle *}\Theta_{\Sigma_{C}} .
\end{align}
The electric current $j$ is non-vanishing:
$
 j := \delta F \ne 0 ,
$
while the magnetic current $k$ is vanishing due to the Bianchi identity and there is no magnetic contribution to the Wilson loop: 
\begin{align}
 k :=  \delta  {}^{\displaystyle *}F 
=  {}^{\displaystyle *}dF 
=  {}^{\displaystyle *}ddA  = 0 
\Longrightarrow  W_C[A]  
 =  \exp \left\{  ie  (N_{\Sigma_{C}},j)   ] \right\} ,
\end{align}
as far as there are no singularities in   $A$.


\begin{figure}[tbp]
\includegraphics[height=2.0cm]{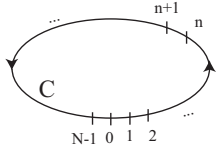}
\caption{
A loop $C$ for the Wilson loop is divided into $N$ infinitesimal segments to obtain the path-integral representation. 
}
\label{fig:W_loop-path-ordering}
\end{figure}

Next, we consider the non-Abelian case.
The non-Abelian Wilson loop operator $W_C[\mathscr{A}]$  (in the representation $R$) is written using the trace and the path ordering as
\begin{equation}
 W_C[\mathscr{A}] :=   {\rm tr}_{R} \left\{ \mathscr{P} \exp \left[ -ig_{{}_{\rm YM}} \oint_C \mathscr{A} \right] \right\}/{\rm tr}_R({\bf 1}) .
\end{equation}
The \textbf{path ordering} $\mathscr{P}$ is defined by  dividing the path $C$ into $N$ infinitesimal segments (See Fig.~\ref{fig:W_loop-path-ordering}):
\begin{align}
 W_C[\mathscr{A}]  = \lim_{N \rightarrow \infty, \epsilon \rightarrow 0}  {\rm tr}_{R} \left\{ \mathscr{P} \prod_{n=0}^{N-1}  \exp \left[ -ig_{{}_{\rm YM}}  \int_{x_n}^{x_{n+1}}   \mathscr{A}  \right]  \right\}/{\rm tr}_R({\bf 1}) .
\end{align}
The troublesome path ordering in the non-Abelian Wilson loop operator can be removed as first shown for $G=SU(2)$ by
[Diakonov and Petrov (1989)]\cite{DP89}, which we call the \textbf{non-Abelian Stokes theorem} (NAST). 
Moreover, the non-Abelian Stokes theorem   for the Lie group $G$ can be obtained as the path-integral representation of the Wilson loop operator
using the coherent state of the Lie group $G$ in an unified way. 
[Kondo (1998), Kondo and Taira (2000), Kondo (2008)]\cite{Kondo98b,KT99,Kondo99Lattice99,Kondo08}.

In order to derive the non-Abelian Stokes theorem,  we follow the standard steps for the path integral: 
\begin{enumerate}
\item
We replace the trace of the operator $\mathscr{O}$ by the integral:
\begin{equation}
 {\rm tr}_{\rm R}(\mathscr{O})/{\rm tr}_R({\bf 1}) 
 = \int d\mu({g}(x_0)) \left< {g}(x_0), \Lambda \right| \mathscr{O}\left| {g}(x_0), \Lambda \right> ,
 \label{C29-trace-integral}
\end{equation}
where 
$d\mu({g})$ is an invariant measure on $G$
and the state is normalized 
$
\left< {g}(x_n), \Lambda | {g}(x_n), \Lambda \right>=1 
 .
 $
\\
\item
We insert a \textbf{complete set} of states at each partition point:
\begin{equation}
 {\bf 1} = \int d\mu({g}(x_n))  \left| {g}(x_n), \Lambda \right> \left< {g}(x_n), \Lambda \right| 
\quad  (n=1, \cdots, N-1) .
\end{equation}
Here the state $\left| {g}, \Lambda \right>$ is constructed by operating a group element ${g} \in G$ to a \textbf{reference state} $\left| \Lambda \right>$ (e.g., the highest-weight state) for a given representation $R$   of the Wilson loop we consider: 
\begin{equation}
 \left| {g} , \Lambda \right> = {g} \left| \Lambda \right> , \quad g \in G .
\end{equation}
\noindent
\item
We take the limit $N \rightarrow \infty$ and $\epsilon \rightarrow 0$ appropriately such that $N\epsilon$ is fixed:
\begin{align}
  W_C[\mathscr{A}]  
=& \lim_{N \rightarrow \infty, \epsilon \rightarrow 0}
\prod_{n=0}^{N-1} \int d\mu({g}(x_n)) \prod_{n=0}^{N-1}
\langle {g}(x_{n+1}), \Lambda | \exp \left[ -ig_{{}_{\rm YM}}  \int_{x_n}^{x_{n+1}}   \mathscr{A}  \right]  | {g}(x_n), \Lambda \rangle  
 .
\end{align}
\end{enumerate}
For taking the limit $\epsilon \rightarrow 0$ in the final step, it is sufficient to retain the $O(\epsilon)$ terms: 
\begin{align}
  & \langle {g}_{n+1}, \Lambda |  \exp \left[ -ig_{{}_{\rm YM}}  \int_{x_n}^{x_{n+1}}   \mathscr{A}  \right] | {g}_{n}, \Lambda \rangle
\nonumber\\
=& \langle  \Lambda |   {g}(x_{n+1})^\dagger  \exp \left[ -ig_{{}_{\rm YM}}  \int_{x_n}^{x_{n+1}}   \mathscr{A}  \right]  g(x_{n}) | \Lambda \rangle
=    \langle  \Lambda | \exp \left[-ig_{{}_{\rm YM}}  \int_{x_n}^{x_{n+1}}  \mathscr{A}^{g}  \right]  | \Lambda \rangle
\nonumber\\
 =&    \langle   \Lambda |   \left[ 1-ig_{{}_{\rm YM}}  \int_{x_n}^{x_{n+1}} d\tau \mathscr{A}^{g}(\tau)   +   O(\epsilon^2) \right] |  \Lambda \rangle
\nonumber\\
 =&   1  -ig_{{}_{\rm YM}}  \int_{x_n}^{x_{n+1}}   \langle  \Lambda |\mathscr{A}^{g}  |  \Lambda \rangle
 +   O(\epsilon^2) 
 \quad (\langle  \Lambda |  \Lambda \rangle = 1)
\nonumber\\
 =&  \exp \left[ -i \epsilon g_{{}_{\rm YM}}  \int_{x_n}^{x_{n+1}}   \langle  \Lambda |\mathscr{A}^{g}  |  \Lambda \rangle  \right] +   O(\epsilon^2) 
  . 
\end{align}
Here $\mathscr{A}^{g}(x)$ agrees with the gauge transformation of $\mathscr{A}(x)$ by the group element $g$: 
\begin{align}
\mathscr{A}^{g}(x)  :=  {g}(x)^\dagger \mathscr{A}(x) {g}(x)
+ ig_{{}_{\rm YM}}^{-1} {g}(x)^\dagger d {g}(x) .
\end{align}
Defining the one-form $A^{g}$ from the  Lie algebra valued one-form $\mathscr{A}^g$ by 
\begin{align}
A^g  :=& \langle \Lambda | \mathscr{A}^{g}      |\Lambda \rangle ,
\
\label{C29-pre-NAST0}
\end{align}
 we arrive at a path-integral representation of the Wilson loop operator (pre-NAST):
\begin{align}
 W_C[\mathscr{A}] 
 =&  \int [d\mu({g})]_C \exp \left( 
-ig_{{}_{\rm YM}}  \oint_C  A^g \right) , 
\quad
[d\mu({g})]_C 
:= \lim_{N \rightarrow \infty, \epsilon \rightarrow 0} \prod_{n=0}^{N-1}  d\mu({g}(x_n)) .
\label{C29-Ag2}
\end{align}
The path-ordering has disappeared. 
\\
Therefore, we can apply the (usual) {Stokes theorem} 
to 
obtain a \textbf{non-Abelian Stokes theorem}:
\begin{equation}
 W_C[\mathscr{A}]  =\int [d\mu(g)]_{\Sigma_{C}}
\exp \left[ -ig_{{}_{\rm YM}} \int_{\Sigma_{C}: \partial \Sigma_{C}=C} F^g  \right] , 
\quad  F^g  =dA^g .
\label{C29-NAST1}
\end{equation}
Here we have replaced the integration measure on the loop $C$ by the integration measure on the surface $\Sigma_{C}$:
\begin{equation}
 [d\mu(g)]_{\Sigma_{C}} :=\prod_{x \in \Sigma_{C}: \partial \Sigma_{C}=C}   d\mu(g(x))  ,
\end{equation}
by inserting additional integration measures,  $1=\int d\mu(g(x))$ for $x \in \Sigma_{C} - C$.
The explicit expression for $F^g$ will be obtained later.

\section{Field decomposition for $SU(2)$ a la  Cho-Duan-Ge-Faddeev-Niemi
}

For the \textbf{highest-weight state} $\left|   {\Lambda} \right> =(\lambda_a)$ of a representation $R$ of a group $G$, we define a matrix $\rho$ with the matrix element $\rho_{ab}$ by
\begin{equation}
 \rho :=  \left|  \Lambda \right> \left< \Lambda \right| 
, \quad 
 \rho_{ab} :=  \left|  \Lambda \right>_a \left< \Lambda \right|_b = \lambda_a \lambda_b^*
 .
\end{equation}
Since  $\left.|\Lambda \right>$ is normalized: 
$\left<   \Lambda |   \Lambda \right>=\lambda_a \lambda_a^*=1$, the trace of $\rho$ has a unity:
\begin{equation}
 {\rm tr}(\rho) = \rho_{aa} 
= 1 
 .
 \label{form11}
\end{equation}
Moreover, the matrix element $\left<  \Lambda \right|   \mathscr{O}  \left|  \Lambda \right>$ of an arbitrary matrix $\mathcal{O}$ is written in the trace form:
\begin{equation}
  \left<  \Lambda \right|   \mathscr{O}  \left|  \Lambda \right> 
 =  {\rm tr}(\rho  \mathscr{O}  )
 ,
 \label{form3}
\end{equation}
since 
$
  \left<  \Lambda \right|   \mathscr{O}  \left|  \Lambda \right>
  = \lambda_b^*  \mathscr{O}_{ba} \lambda_a 
  = \rho_{ab}  \mathscr{O}_{ba}   
 = {\rm tr}(\rho  \mathscr{O}  )
 .
$

By using the operator 
$\rho$, 
 the ``Abelian'' field $A^g$ 
is written in the trace form of a matrix: 
\begin{align}
   A^g(x) =& 
 \langle \Lambda | \mathscr{A}^g (x)    |\Lambda \rangle  
=  {\rm tr}\{\rho \mathscr{A}^g(x)   \} 
=  {\rm tr}\{ g(x) \rho  g^\dagger(x)  \mathscr{A}(x) \} + ig_{{}_{\rm YM}}^{-1} {\rm tr}\{ \rho  g^\dagger(x)  d   g(x)  \} 
 .
 \label{C29-def-Ag}
\end{align}
By introducing the traceless field $\tilde{\bm{n}}(x)$ defined by [which we call the \textbf{color (direction) field} after the normalization]
\begin{equation}
 \tilde{\bm{n}}(x) :=  g(x) \left[ \rho - \frac{\bm{1}}{{\rm tr}(\bm{1})} \right]  g^\dagger(x) 
= g(x)  \rho g^\dagger(x) - \frac{\bm{1}}{{\rm tr}(\bm{1})}    ,
  \label{C29-tilde-n}
\end{equation}
the ``Abelian'' field $A^g$   is rewritten as
\begin{align}
   A^g_\mu(x) =& {\rm tr}\{ \tilde{\bm{n}}(x)  \mathscr{A}_\mu(x) \} + ig_{{}_{\rm YM}}^{-1} {\rm tr}\{ \rho  g^\dagger(x)  \partial_\mu   g(x)  \} 
 .
 \label{C29-def-Ag2}
\end{align}

We proceed to perform the \textbf{decomposition of the  Yang-Mills field} $\mathscr{A}_\mu(x)$  into two pieces:
\begin{equation}
 \mathscr{A}_\mu(x)
=\mathscr{V}_\mu(x)+\mathscr{X}_\mu(x)
 .
\end{equation}
We simply require that $\mathscr{X}_\mu(x)$ satisfies the condition:[which we call the second \textbf{defining equation}]
\begin{align}
\text{(ii)}  \quad  \mathscr{X}_\mu(x) \cdot \bm{n}(x) = 2{\rm tr}\{  \mathscr{X}_\mu(x) \bm{n}(x) \} = 0.
 \label{defeq-X-1}
\end{align}
Then $\mathscr{X}_\mu(x)$ disappears from the Wilson loop operator, since $A^g_\mu(x)$ is written without $\mathscr{X}_\mu(x)$:
\begin{align}
   A^g_\mu(x) =& {\rm tr}\{ \tilde{\bm{n}}(x)  \mathscr{V}_\mu(x) \} + ig_{{}_{\rm YM}}^{-1} {\rm tr}\{ \rho  g^\dagger(x)  \partial_\mu   g(x)  \} 
 .
\end{align}
Consequently, the Wilson loop operator $W_C[\mathscr{A}]$ 
can be reproduced by the \textbf{restricted field} variable   $\mathscr{V}_\mu(x)$ alone. 
This is called the \textbf{restricted field dominance} for the Wilson loop operator.
For arbitrary loop $C$ and any representation $R$, the Wilson loop operators satisfies
\begin{equation}
\text{(a)}  \quad W_C[\mathscr{A}]= W_C[\mathscr{V}]
 .
 \label{C29-W-dominant0}
\end{equation} 
This does not necessarily imply the restricted field dominance for the Wilson loop average: 
\begin{equation}
\langle W_C[\mathscr{A}] \rangle_{\rm YM}
= \langle W_C[\mathscr{V}] \rangle_{\rm YM} ,
\end{equation} 
which holds only when the cross term between $\mathscr{V}$ and $\mathscr{X}$ in the action can be neglected. 

We look for the \textbf{gauge covariant decomposition}  which  means that the decomposition holds after the gauge transformation:
\begin{equation}
 \mathscr{A}_\mu^\prime(x)
=\mathscr{V}_\mu^\prime(x)+\mathscr{X}_\mu^\prime(x)
 .
\end{equation}
For the condition (ii) [eq.(\ref{defeq-X-1})] to be gauge covariant, 
the transformation of the color field $\bm{n}$ given by
\begin{equation}
 g(x) \to U(x) g(x) \Longrightarrow  \bm{n}(x) \to \bm{n}^\prime(x) = U(x)  \bm{n}(x) U^{\dagger}(x) .
\end{equation}
requires that $\mathscr{X}_\mu(x)$ transforms as  an adjoint (matter) field:
\begin{align}
  \mathscr{X}_\mu(x) & \rightarrow \mathscr{X}_\mu^\prime(x) = U(x)  \mathscr{X}_\mu(x) U^{\dagger}(x) 
 \label{C29-X-ctransf}
 , 
\end{align}
This immediately means that $\mathscr{V}_\mu(x)$ must transform   just like the original gauge field $\mathscr{A}_\mu(x)$:
\begin{align}
  \mathscr{V}_\mu(x) & \rightarrow \mathscr{V}_\mu^\prime(x) = U(x)  \mathscr{V}_\mu(x) U^{\dagger}(x) + ig_{{}_{\rm YM}}^{-1} U(x)  \partial_\mu U^{\dagger}(x) 
 ,
 \label{C29-V-ctransf}
\end{align}
since 
$
  \mathscr{A}_\mu(x)   \rightarrow \mathscr{A}_\mu^\prime(x) = U(x)  \mathscr{A}_\mu(x) U^{\dagger}(x) + ig_{{}_{\rm YM}}^{-1} U(x)  \partial_\mu U^{\dagger}(x) 
 .
$

These transformation properties impose restrictions on the requirement to be imposed on the restricted field $\mathscr{V}_\mu(x)$. 
Such a candidate is [covariant constantness of the color field] [which we call the first \textbf{defining equation}]:
\begin{align}
 \text{(I)}  \quad  \mathscr{D}_\mu[\mathscr{V}] \bm{n} = 0 
\quad
(\mathscr{D}_\mu[\mathscr{V}]  :=   \partial_\mu   -ig_{{}_{\rm YM}} [ \mathscr{V}_\mu,  \cdot ]  ) , 
 \label{defeq-V-1}
\end{align}
since the covariant derivative transforms in the adjoint way:
$\mathscr{D}_\mu[\mathscr{V}(x)] \to U(x)(\mathscr{D}_\mu[\mathscr{V}](x))U^\dagger(x)$. 

For $G=SU(2)$, it is  shown that the two conditions (I) and (ii), i.e., (\ref{defeq-V-1}) and (\ref{defeq-X-1})  [the \textbf{defining equations} for the decomposition] are compatible and determine the decomposition uniquely: 
\begin{align}
 \mathscr{A}_\mu(x)= \mathscr{V}_\mu(x) +& \mathscr{X}_\mu(x),
  \nonumber
\\
   \mathscr{V}_\mu(x) =& c_\mu(x)\bm{n}(x) + ig_{{}_{\rm YM}}^{-1} [\bm{n}(x) , \partial_\mu \bm{n}(x) ] , 
\quad
   c_\mu(x) :=   \mathscr{A}_\mu(x)  \cdot \bm{n}(x) ,
  \nonumber
\\
  \mathscr{X}_\mu(x) =&  -ig_{{}_{\rm YM}}^{-1} [ \bm{n}(x) ,  \mathscr{D}_\mu[\mathscr{A}] \bm{n}(x) ] .
\end{align}
This is the same as the \textbf{Cho--Duan-Ge (CDG) decomposition} or \textbf{Cho--Duan-Ge--Faddeev-Niemi (CDGFN) decomposition}  
[Cho(1980), Duan-Ge (1979), Faddeev-Niemi (1998)] 
\cite{Cho80,DG79,FN99,Shabanov99}.

The condition (I)  means that 
the field strength $\mathscr{F}_{\mu\nu}^{[\mathscr{V}]}(x)$ of the field $\mathscr{V}_\mu(x)$ and $\bm{n}(x)$ commute:
\begin{equation}
[ \mathscr{F}_{\mu\nu}^{[\mathscr{V}]}(x) , \bm{n}(x) ]
= 0   
 .
 \label{C29-F-m}
\end{equation}
This follows from the identity: 
\begin{equation}
 [ \mathscr{F}_{\mu\nu}^{[\mathscr{V}]}, \bm{n} ]
 =    ig_{{}_{\rm YM}}^{-1} [ \mathscr{D}_\mu^{[\mathscr{V}]},  \mathscr{D}_\nu^{[\mathscr{V}]} ] \bm{n}  
 ,
\end{equation}
which is derived from
\begin{equation}
 \mathscr{F}_{\mu\nu}^{[\mathscr{V}]} 
= ig_{{}_{\rm YM}}^{-1} [ \mathscr{D}_\mu^{[\mathscr{V}]}, \mathscr{D}_\nu^{[\mathscr{V}]} ] ,
\quad
 \mathscr{D}_\mu^{[\mathscr{V}]} := \partial_\mu -ig_{{}_{\rm YM}} [ \mathscr{V}_\mu , \cdot ] .
\end{equation}
For $SU(2)$, (\ref{C29-F-m}) means that $\mathscr{F}_{\mu\nu}^{[\mathscr{V}]}(x)$ is proportional to $\bm{n}(x)$: 
\begin{equation}
 \mathscr{F}_{\mu\nu}^{[\mathscr{V}]}(x) =   f_{\mu\nu}(x) \bm{n}(x) \Longrightarrow
 f_{\mu\nu}(x) = \bm{n}(x)  \cdot \mathscr{F}_{\mu\nu}^{[\mathscr{V}]}(x) = 2{\rm tr}[\bm{n}(x) \mathscr{F}_{\mu\nu}^{[\mathscr{V}]}(x)]
    ,
    \label{C29-F=Fn}
\end{equation}
since $\mathscr{F}_{\mu\nu}^{[\mathscr{V}]}(x)$ is traceless and cannot have a part proportional to the unit matrix.  

\section{Field decomposition for $SU(N)$: new options}

For $G=SU(N)$ ($N \ge 3$), (I) and (ii) are not sufficient to uniquely determine the decomposition.  The condition (ii) [eq.(\ref{defeq-X-1})] must be modified: 
[Kondo, Shinohara and Murakami (2008)]\cite{KSM08}
\\
(II)  $\mathscr{X}^\mu(x)$  does not have the $\tilde{H}$-commutative part, i.e., $\mathscr{X}^\mu(x)_{\tilde{H}}=0$:
\begin{align}
\text{(II)}  \quad  & 0 =  \mathscr{X}^\mu(x)_{\tilde{H}} := \mathscr{X}^\mu(x)  - \frac{2(N-1)}{N}  [\bm{n}(x) , [\bm{n}(x) , \mathscr{X}^\mu(x) ]] 
\nonumber\\
& \Longleftrightarrow \mathscr{X}^\mu(x)  =  \frac{2(N-1)}{N}  [\bm{n}(x) , [\bm{n}(x) , \mathscr{X}^\mu(x)  ]]
\label{C29-defXL2b}
 . 
\end{align}
This condition is also gauge covariant. 
Note that the condition (ii)[eq.(\ref{defeq-X-1})] follows from (II)[eq.(\ref{C29-defXL2b})]. 
For $G=SU(2)$, i.e., $N=2$, the condition (II)[eq.(\ref{C29-defXL2b})] reduces to (ii)[eq.(\ref{defeq-X-1})].
By solving (I)[eq.(\ref{defeq-V-1})] and (II)[eq.(\ref{C29-defXL2b})], 
$\mathscr{X}_\mu(x)$ is determined as
\begin{align}
 \mathscr{X}_\mu(x)   
=& -ig_{{}_{\rm YM}}^{-1}  \frac{2(N-1)}{N}  [\bm{n}(x), \mathscr{D}_\mu[\mathscr{A}]\bm{n}(x) ]
 \in Lie(G/\tilde H)
 ,
\label{C29-X-def1}
\\
\mathscr V_\mu(x) 
 =& \mathscr C_\mu(x) 
  +\mathscr B_\mu(x) \in \mathscr{L}ie(G)
 ,
\nonumber\\
 & \mathscr{C}_\mu(x) 
=  \mathscr{A}_\mu(x)  - \frac{2(N-1)}{N}   [\bm{n}(x) , [ \bm{n}(x) , \mathscr{A}_\mu(x) ] ]
\in  \mathscr{L}ie(\tilde{H}) 
,
\nonumber\\
& \mathscr{B}_\mu(x) 
=  i g_{{}_{\rm YM}}^{-1} \frac{2(N-1)}{N}[\bm{n}(x) , \partial_\mu  \bm{n}(x)  ] 
\in \mathscr{L}ie(G/\tilde{H}) 
 .
 \label{NLCV-minimal}
\end{align}

\section{Non-Abelian Stokes theorem (2)}
\small 

Finally, we can show that the field strength $F_{\mu\nu}^g:=  \partial_\mu A_\nu - \partial_\nu A_\mu$  in NAST (\ref{C29-NAST1}) is cast into the form:
\begin{align}
 F_{\mu\nu}^g(x) 
=   \sqrt{\frac{2(N-1)}{N}} &
 { {\rm tr}\{ \bm{n}(x) \mathscr{F}_{\mu\nu} [\mathscr{V}](x) \} }
 + ig_{{}_{\rm YM}}^{-1} {\rm tr}\{ \rho  g^\dagger(x)[\partial_\mu , \partial_\nu] g(x)   \} 
 .
\\
 { {\rm tr}\{ \bm{n}(x) \mathscr{F}_{\mu\nu} [\mathscr{V}](x) \} }
  =&   \partial_\mu {\rm tr}\{ \bm{n}(x)  \mathscr{V}_\nu(x) \}
-  \partial_\nu {\rm tr}\{ \bm{n}(x)  \mathscr{V}_\mu(x) \}
   + \frac{2(N-1)}{N} ig_{{}_{\rm YM}}^{-1}  {\rm tr}\{ \bm{n}(x) [ \partial_\mu \bm{n}(x) , \partial_\nu \bm{n}(x) ] \} 
 ,
\end{align}
where the normalized and traceless \textbf{color direction field} $\bm{n}(x)$ is defined by
\begin{equation}
 \bm{n}(x) 
 = \sqrt{\frac{N}{2(N-1)}} g(x) \left[ \rho - \frac{\bm{1}}{{\rm tr}(\bm{1})} \right]  g^\dagger(x)  , \quad
 g(x) \in G 
 .
 \label{C29-n-def}
\end{equation}
Thus the Wilson loop operator can be rewritten   in terms of new variables:
\begin{align}
 W_C[\mathscr{A}] =& \int [d\mu(g)]_{\Sigma_{C}} 
 \exp \Big[  -ig_{{}_{\rm YM}} \frac12 \sqrt{\frac{2(N-1)}{N}}
 \int_{\Sigma_{C}: \partial \Sigma_{C}=C} 2{\rm tr} \{ \bm{n}  \mathscr{F}[\mathscr{V}]  \} \Big] .
\end{align}
Incidentally, the last part $ig_{{}_{\rm YM}}^{-1} {\rm tr} \{ \rho g(x)^\dagger [\partial_\mu, \partial_\nu] g(x) \}$ in $F_{\mu\nu}^g(x)$ corresponds  to the \textbf{Dirac string}. 
This term is not gauge invariant and does not contribute to the Wilson loop operator in the end, since it disappears after the group integration $d\mu(g)$ is performed. 
 
In this way we obtain another expression of the NAST for the Wilson loop operator: For $SU(N)$ in the \textbf{fundamental representation}:
\begin{equation}
 W_C[\mathscr{A}] 
= \int [d\mu({g})]  \exp \left\{ -ig_{{}_{\rm YM}}  \frac12  \sqrt{\frac{2(N-1)}{N}} [ (  N_{\Sigma_{C}},j  ) + (  \Xi_{\Sigma_{C}}, k)   ] \right\} ,
\end{equation}
where 
we have defined the $(D-3)$-form $k$ and one-form $j$ by
\begin{align}
k:=  \delta {}^{\displaystyle *}f  , \quad  j:=  \delta f  , \quad
f  := 2{\rm tr}\{ \bm{n} \mathscr{F}  [\mathscr{V}]\} ,
\end{align}
and
we have defined the $(D-3)$-form $\Xi_{\Sigma_{C}}$ and one-form $N_{\Sigma_{C}}$  by 
($\Xi_{\Sigma_{C}}$ is the $D$-dim. solid angle)
\begin{equation}
 \Xi_{\Sigma_{C}} :=  {}^{\displaystyle *} d \Delta^{-1} \Theta_{\Sigma_{C}}  = \delta  \Delta^{-1}  {}^{\displaystyle *}\Theta_{\Sigma_{C}}  , \quad
 N_{\Sigma_{C}} := \delta  \Delta^{-1}  \Theta_{\Sigma_{C}} ,
\end{equation}
with the inner product for the two forms defined by
\begin{align}
 & ( \Xi_{\Sigma_{C}},k) 
= \frac{1}{(D-3)!} \int d^Dx k^{\mu_1 \cdots \mu_{D-3}}(x) \Xi^{\mu_1 \cdots \mu_{D-3}}_{\Sigma_{C}}(x) ,
\quad   (N_{\Sigma_{C}},j) =  \int d^Dx j^{\mu}(x) N^{\mu}_{\Sigma_{C}}(x) .
\end{align}
Thus the  Wilson loop operator can  be expressed by the electric current $j$ and the  monopole current $k$.
\\
The magnetic monopole described by the current $k$ is a topological object of \textbf{co-dimension} 3:
\begin{itemize}
\item
$D=3$:  $0$-dimensional point defect $\rightarrow$ point-like magnetic monopole (cf. Wu-Yang type)
\item
$D=4$: $1$-dimensional line defect  $\rightarrow$ {  magnetic monopole loop (closed loop)}
\end{itemize}

\noindent
\subsection{$SU(2)$ case}
For $SU(2)$, the gauge-invariant magnetic-monopole current $(D-3)$-form $k$ is obtained  
\begin{align}
k= 
\delta {}^*f , \ 
  f_{\mu\nu} 
= 2{\rm tr}\{ \bm{n} \mathscr{F}_{\mu\nu} [\mathscr{V}]\}
=  \partial_\mu 2{\rm tr} \{ \bm{n}  \mathscr{A}_\nu  \} - \partial_\nu 2{\rm tr} \{ \bm{n}  \mathscr{A}_\mu  \} 
+ ig_{{}_{\rm YM}}^{-1} 2{\rm tr} \{ \bm{n} [\partial_\mu \bm{n} , \partial_\nu \bm{n}  ] \} 
 .
\end{align}
For the fundamental representation of $SU(2)$, the highest-weight state 
$
| \Lambda \rangle 
$ 
yields the color field:
\begin{align}
& | \Lambda \rangle 
=  
\begin{pmatrix}
 1 \\
 0 
\end{pmatrix}
\Longrightarrow 
\rho :=  | \Lambda \rangle \langle \Lambda |
= 
\begin{pmatrix}
 1 \\
 0 
\end{pmatrix}
 (1,0)
= 
\begin{pmatrix}
 1 & 0 \\
 0 & 0 
\end{pmatrix}
\Longrightarrow 
 \rho - \frac12 \mathbf{1} = \frac{\sigma_3}{2} ,
\nonumber\\
& \Longrightarrow \bm{n}(x) 
=   g(x) \frac{\sigma_3}{2} g(x)^\dagger 
\in SU(2)/U(1) \simeq S^2 \simeq  P^1(\mathbb{C}) .
\end{align}
The magnetic charge $q_m$ obeys the \textbf{quantization condition} a la Dirac:
\begin{equation}
 q_m := \int d^3x k^0 = 4\pi  g_{{}_{\rm YM}}^{-1} \ell , \ \ell \in \mathbb{Z} .
\end{equation}
This is suggested from a nontrivial Homotopy group of the map $\bm{n}: S^2 \rightarrow SU(2)/U(1)$:  
\begin{equation}
 \pi_2(SU(2)/U(1))=\pi_1(U(1))=\mathbb{Z} .
\end{equation}
This should be compared with 
the Abelian magnetic monopole due to 't Hooft-Polyakov   associated with the spontaneous symmetry breaking $G=SU(2) \to H=U(1)$:
\begin{equation}
  \bm{n}^A \leftrightarrow \hat{\phi}^A(x)/|\hat{\phi}(x)| .
\end{equation}

\noindent
\subsection{$SU(3)$ case}
For $SU(3)$, the gauge--invariant magnetic--monopole current $(D-3)$-form $k$ is given by [Kondo (2008)]\cite{Kondo08}
\begin{align}
k=  
 \delta {}^*f,
\quad
 f_{\mu\nu}  
  :=   \partial_\mu 2{\rm tr} \{ \bm{n}  \mathscr{A}_\nu  \} - \partial_\nu 2{\rm tr} \{ \bm{n}  \mathscr{A}_\mu  \} 
+ \frac{4}{3} ig_{{}_{\rm YM}}^{-1} 2{\rm tr} \{ \bm{n} [\partial_\mu \bm{n} , \partial_\nu \bm{n}  ] \} 
 .
\end{align}
For the fundamental representation of $SU(3)$, the highest-weight state $| \Lambda \rangle$ yields the color field:
\begin{align}
& | \Lambda \rangle 
=  
\begin{pmatrix}
 1 \\
 0 \\
 0 
\end{pmatrix} 
\Longrightarrow 
\rho :=  | \Lambda \rangle \langle \Lambda |
= 
\begin{pmatrix}
 1 & 0 & 0 \\
 0 & 0 & 0 \\
 0 & 0 & 0 \\
\end{pmatrix} 
\Longrightarrow 
 \rho - \frac13 \mathbf{1} 
= \frac{-1}{3} 
\begin{pmatrix}
 -2 & 0 & 0 \\
 0 & 1 & 0 \\
 0 & 0 & 1 \\
\end{pmatrix} 
,
\\
& \Longrightarrow \bm{n}(x) 
=   {g(x)}
\frac{-1}{2\sqrt{3}} 
\begin{pmatrix}
 -2 & 0 & 0 \\
 0 &  {1} & 0 \\
 0 & 0 &  {1} \\
\end{pmatrix} 
 {g(x)^\dagger} 
\in SU(3)/U(2) \simeq P^2(\mathbb{C}) . 
\end{align}
The matrix ${\rm diag.}(-2,1,1)$ is degenerate. Using the Weyl symmetry (discrete global symmetry as a subgroup of color symmetry), it is changed into $\lambda_8$. 
This color field describes a \textbf{non-Abelian magnetic monopole}, which corresponds to the spontaneous symmetry breaking $SU(3) \to U(2)$ in the gauge-Higgs model. 

The  magnetic charge obeys the quantization condition: 
\begin{equation}
 q_m' := \int d^3x k^0 = 2\pi \sqrt{3} g_{{}_{\rm YM}}^{-1} n' , \ n' \in \mathbb{Z} .
\end{equation}
This is suggested from a nontrivial Homotopy group of the map   $\bm{n}: S^2 \rightarrow SU(3)/U(2)$  
\begin{equation}
   \pi_2(SU(3)/[SU(2) \times U(1)])=\pi_1(SU(2) \times U(1)) 
=\pi_1(U(1))=\mathbb{Z} .
\end{equation}

For a \textbf{reference state} $|\Lambda \rangle$ of a given representation of a Lie group $G$,
the \textbf{maximal stability subgroup} $\tilde H$ is defined to be a subgroup leaving $|\Lambda \rangle$ invariant (up to a phase $\phi(h)$):
\begin{equation}
  h \in \tilde H \Longleftrightarrow h |\Lambda \rangle = |\Lambda \rangle e^{i\phi(h)} .
\end{equation}
Then a group element $g$ of $G$ is decomposed as
\begin{equation}
   g = \xi h  \in G, \quad \xi \in G/\tilde H, \quad h \in \tilde H .
\end{equation}
Therefore, we have
\begin{equation}
  |g,\Lambda \rangle :=g |\Lambda \rangle = \xi h  |\Lambda \rangle
  = \xi  |\Lambda \rangle  e^{i\phi(h)}
= |\xi , \Lambda \rangle  e^{i\phi(h)}
 .
\end{equation}

\begin{figure}[ptb]
\includegraphics[scale=0.35]{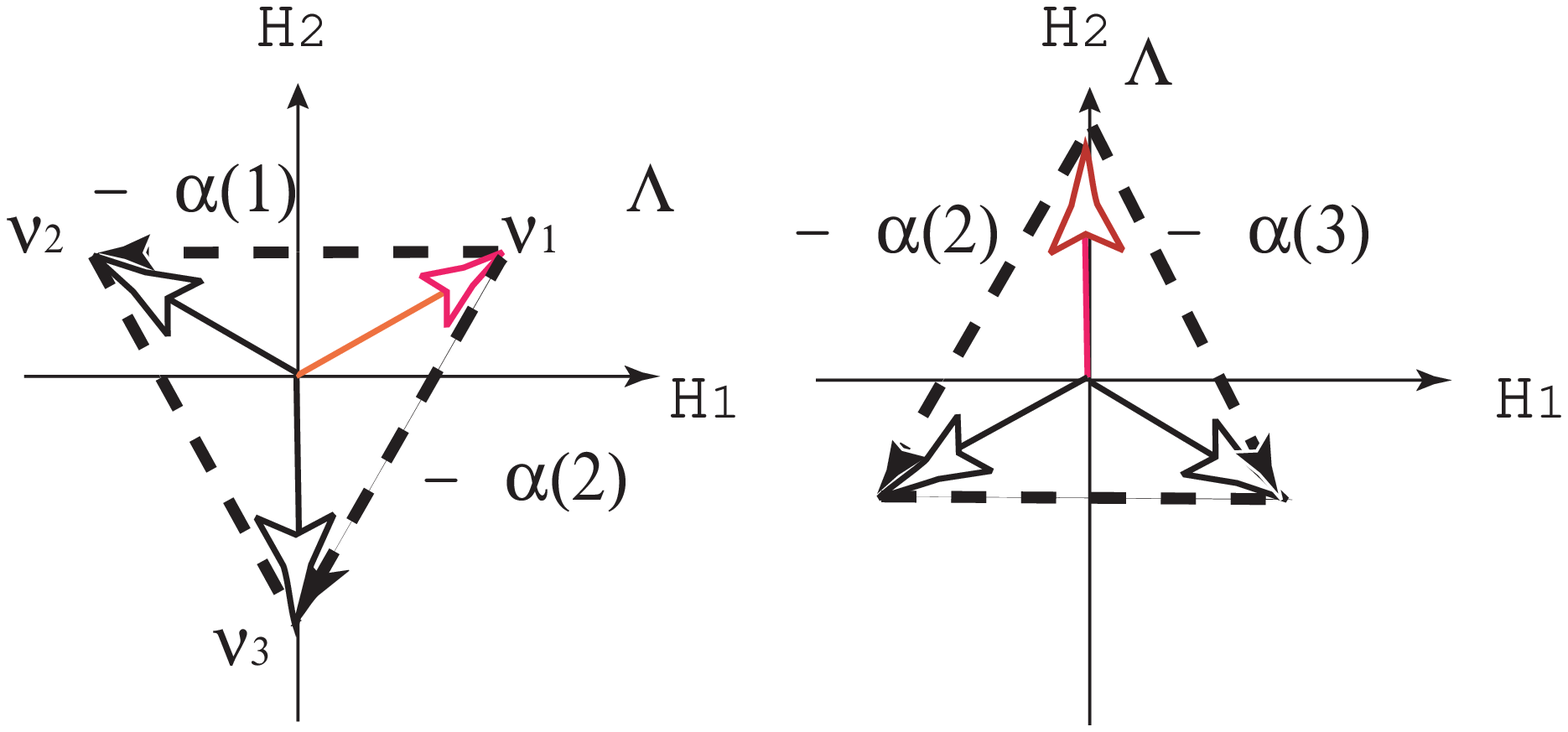}
\includegraphics[scale=0.25]{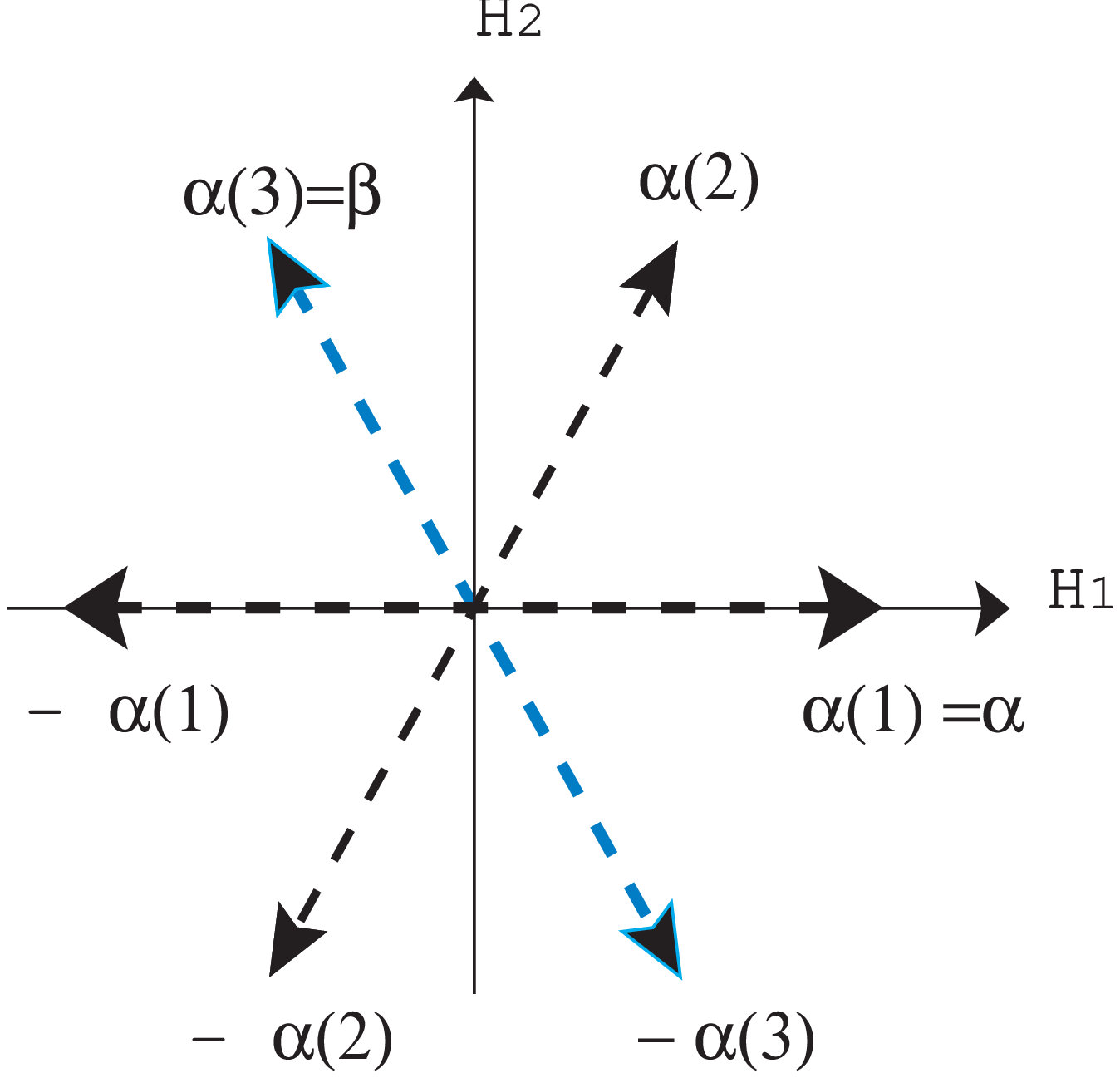}
\caption{
The relationships among the weight vectors $\vec{\nu}_1, \vec{\nu}_2, \vec{\nu}_3$ in the fundamental representations ${\bf 3}$     and the root vectors $\vec{\alpha}^{(1)}, \vec{\alpha}^{(2)}, \vec{\alpha}^{(3)}$  in $SU(3)$.
We find
$
\vec{\nu}_1  \perp  \vec{\alpha}^{(3)}, - \vec{\alpha}^{(3)}
$.
Here   $\vec\Lambda  =\vec{\nu}_1:=(\frac{1}{2},\frac{1}{2\sqrt{3}})$ is the highest weight of the fundamental representation ${\bf 3}$.
}
 \label{C28-fig:fundamental-weight3}
\end{figure}

Every  representation  $R$ of $SU(3)$ which is specified by the Dynkin index [m,n] belongs to  (I) or (II):

\noindent 
\begin{enumerate}
\item
[(I)] [Maximal case] 
$
m \ne 0 \ \text{and} \ n \ne 0   \Longrightarrow \tilde H = H= U(1) \times U(1) .
$ maximal torus
\\
e.g., adjoint rep.[1,1], $\{ H_1, H_2  \} \in u(1)+u(1)$,

\item
[(II)] [Minimal case] 
$
m=0 \ \text{or} \ n= 0   \Longrightarrow \tilde H = U(2) .
$
\\
This case occurs when  {the weight vector $\Lambda$} is orthogonal to    {some of the root vectors}. 
(See Fig.~\ref{C28-fig:fundamental-weight3})
\\
e.g., fundamental rep. [1,0], 
$\{ H_1, H_2, E_\beta, E_{-\beta} \} \in u(2)$,
where $\Lambda \perp \beta, -\beta$. 
\end{enumerate}

\noindent
\section{Reformulating Yang-Mills theory using new variables}

\begin{figure}[tbp]
\includegraphics[scale=0.4]{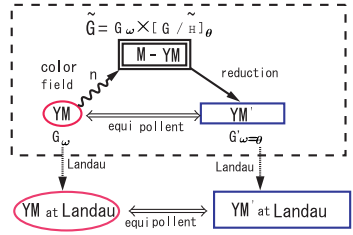}
\caption{\small  \cite{KSM08}
The relationship between the original Yang-Mills (YM) theory and the reformulated Yang-Mills (YM') theory.  A single color field $\bm{n}$ is introduced to enlarge the original Yang-Mills theory with a gauge group $G$ into the master Yang-Mills (M-YM) theory  with the enlarged gauge symmetry $\tilde{G}=G \times G/\tilde{H}$.  The reduction conditions are imposed to reduce the master Yang-Mills theory to the reformulated Yang-Mills theory with the equipollent gauge symmetry $G^\prime$. 
In addition, we can impose any over-all gauge fixing condition, e.g., Landau gauge to both the original YM theory and the reformulated YM' theory. 
}
\label{R05-fig:enlarged-YM}
\end{figure}

We consider the change of variables from   $\mathscr A_\mu$ to new field variables $\mathscr C_\mu$, $\mathscr X_\mu$ and $\bm n$: (See  [Kondo, Murakami and Shinohara (2005)] for $SU(2)$, and   [Kondo, Shinohara and Murakami (2008)] for $SU(N)$ \cite{KMS05,KMS06,KSM08})
\begin{equation}
 \mathscr A_\mu^A  \Longrightarrow (\bm n^\beta, \mathscr C_\mu^k  ,  \mathscr X_\mu^b )  
 ,
\end{equation} 
\begin{itemize}
\item
$\mathscr A_\mu \in Lie(G)
\rightarrow \#[\mathscr A_\mu^A]=D  \cdot  {\rm dim}G=D(N^2-1)$

\item
$\mathscr C_\mu \in Lie(\tilde{H})=u(N-1) \rightarrow  \#[\mathscr C_\mu^k]=D  \cdot  {\rm dim}\tilde{H}=D(N-1)^2 
$

\item
$\mathscr X_\mu \in Lie(G/\tilde{H})
\rightarrow \#[\mathscr X_\mu^b]
=D  \cdot  {\rm dim}(G/\tilde{H})= D(2N-2)   
$

\item
$\bm{n}  \in Lie(G/\tilde{H})
\rightarrow \#[\bm{n}^\beta]= {\rm dim}(G/\tilde{H})
=2(N-1)$ .

\end{itemize}
The new theory written in terms of new variables $(\bm n^\beta, \mathscr C_\mu^k ,  \mathscr X_\mu^b )$ has the $2(N-1)$ extra degrees of freedom.  
Therefore, we must give a procedure for eliminating the $2(N-1)$ extra degrees of freedom to obtain the new theory  which is equipollent to the original one. For this purpose, we impose $2(N-1)$ constraints $\bm\chi=0$, which we call the \textbf{reduction condition} (See Fig.~\ref{R05-fig:enlarged-YM}):
\begin{itemize}
\item
$
\bm\chi \in Lie(G/\tilde{H}) \rightarrow \#[\bm\chi^a]= {\rm dim}(G/\tilde{H}) 
=2(N-1)=\#[\bm{n}^\beta]
$.
 
\end{itemize}

A choice of the reduction condition in the minimal option   is to minimize the functional $F_{\rm red}[\mathscr A, \bm n]$:
\begin{align}
  \delta F_{\rm red}[\mathscr A, \bm n] = \int d^Dx  \frac12 g^2  \mathscr  X_\mu\cdot \mathscr  X^\mu
  =  \frac{2(N-1)^2}{N^2}\int d^Dx(\bm n \times D_\mu[\mathscr  A]\bm n)^2
  =  \frac{N-1}{N}\int d^Dx(D_\mu[\mathscr  A]\bm n)^2,
\end{align}
with respect to the enlarged gauge transformation: 
\begin{align}
& \delta\mathscr  A_\mu=D_\mu[\mathscr  A]\bm\omega 
\quad 
(\bm\omega \in \mathscr{L}ie(G) ) ,
\nonumber\\
 & \delta\bm n
 =ig[\bm n , \bm\theta ]
 =ig[\bm n , \bm\theta_\perp ]
\quad
 (\bm\theta_\perp \in \mathscr{L}ie(G/\tilde{H})) .
\end{align}
In fact, the enlarged gauge transformation of the functional $F_{\rm red}[\mathscr A, \bm n]$ is
\begin{align}
 \delta F_{\rm red}[\mathscr A, \bm n]
 = \delta   \int d^Dx \frac12 (D_\mu[\mathscr A]\bm n)^2
=g    \int d^Dx
   (\bm\theta_\perp-\bm\omega_\perp)
   \cdot i[\bm n, D^\mu[\mathscr  A]D_\mu[\mathscr  A]\bm n ]
 ,
\end{align}
where $\bm\omega_\perp$ denotes the component of $\bm\omega$ in the direction $\mathscr{L}(G/\tilde{H})$. 
\\
For $\bm\omega_\perp = \bm\theta_\perp$ (diagonal part of $G \times G/\tilde{H}$)  $\delta  F_{\rm red}[\mathscr A, \bm n]=0$ imposes no condition, 
while  
\\
for $\bm\omega_\perp   \not=  \bm\theta_\perp$ (off-diagonal part of $G \times G/\tilde{H}$) it implies the constraint: 
\begin{equation}
\bm\chi[\mathscr A,\bm n]
 := [\bm n , D^\mu[\mathscr A]D_\mu[\mathscr A]\bm n ]
 \equiv0 
  .
\label{eq:nMAG_minimal_differential}
\end{equation}
Note that the number of constraint is 
$\#[\bm\chi]= {\rm dim}(G \times G/\tilde{H})- {\rm dim}(G)= {\rm dim}(G/\tilde{H})$ as desired. 
\\
Finally, we have an equipollent Yang-Mills theory with  \textbf{the residual local gauge symmetry} 
$G':=SU(N)^{\rm local}_{\omega^\prime}$ with the gauge transformation parameter: 
\begin{equation}
\bm\omega^\prime(x)
=(\bm\omega_\parallel(x),\bm\omega_\perp(x) )=(\bm\omega_\parallel(x), \bm\theta_\perp(x)) , 
\quad  \bm\omega_\perp(x)=\bm\theta_\perp(x) .
\end{equation}

\begin{table}[ptb]
\caption{}
\label{}
 \begin{tabular}{l|cl}
    &  original YM & $\Longrightarrow$ reformulated YM \\ \hline
  field variables & $\mathscr A_\mu^A \in \mathscr{L}(G)$ & $\Longrightarrow$ $\bm n^\beta, \mathscr C_\nu^k  ,  \mathscr X_\nu^b$ \\ 
  action & $S_{\rm YM}[\mathscr A]$ & $\Longrightarrow$ $\tilde S_{\rm YM}[\bm n, \mathscr{C},\mathscr{X}]$ \\
  integration measure & $\mathcal{D}\mathscr{A}_\mu^A$ 
& $\Longrightarrow$  $\mathcal{D}n^\beta \mathcal{D}\mathscr{C}_\nu^k   \mathcal{D}\mathscr{X}_\nu^b
\tilde{J}  \delta( {\tilde{\bm\chi}}) 
  \Delta_{\rm FP}^{\rm red}[\bm n,c, \mathscr X]$ \\
 \end{tabular}
\end{table}

\noindent
At the same time, the color field 
\begin{equation}
\bm{n}(x) \in \mathscr{L}ie(G/\tilde{H})
\end{equation}
must be obtained by solving the  {reduction condition $\bm\chi=0$} for a given $\mathscr{A}$, e.g., 
\begin{equation}
 {\bm\chi[\mathscr{A},\bm{n}]}
 :=[ \bm{n} ,  D^\mu[\mathscr{A}]D_\mu[\mathscr{A}]\bm{n} ]
   \in \mathscr{L}ie(G/\tilde{H})
  .
\label{eq:diff-red}
\end{equation}
Here $\tilde{\bm\chi}=0$ is the reduction condition written in terms of the new  variables:
\begin{equation}
\tilde{\bm\chi} 
 :=\tilde{\bm\chi} [\bm n, \mathscr{C},\mathscr{X}]
 :=D^\mu[\mathscr{V}]\mathscr{X}_\mu 
 , 
\end{equation}
and $\Delta_{\rm FP}^{\rm red}$ is  the Faddeev-Popov determinant associated with the reduction condition:
\begin{equation}
 \Delta_{\rm FP}^{\rm red}
:= \det\left(\frac{\delta\bm\chi}{\delta{\bm\theta}}\right)_{\bm{\chi}=0}
=   \det\left(\frac{\delta\bm\chi}{\delta\bm n^\theta}\right)_{\bm{\chi}=0} ,
\end{equation}
which is obtained by  the BRST method as
$
  \Delta_{\rm FP}^{\rm red}[\bm n,c,\mathscr X] 
= \det \{-D_\mu[\mathscr V+\mathscr X]D_\mu[\mathscr V-\mathscr  X]\}  .
$
\\
The Jacobian  $\tilde{J}$  is  very simple, irrespective of the choice of the reduction condition: \cite{Kondo06,KKSS14}
\begin{equation}
  \tilde{J} = 1 .
\end{equation}

Thus the  Wilson loop average in the original theory defined by
\begin{align}
 W(C) :=  \langle W_C[\mathscr{A}]  \rangle_{\rm YM}
  = Z_{{\rm YM}}^{-1} \int \mathcal{D} \mathscr{A} e^{-S_{\rm YM}[\mathscr{A}] }  W_C[\mathscr{A}] 
  ,
\end{align}
is defined in the reformulated Yang-Mills  theory:
\begin{align}
  \langle W_C[\mathscr{A}]  \rangle_{\rm YM^\prime}
  =& Z_{{\rm YM}^\prime}^{-1}  \int  [d\mu(g)]  \int \mathcal{D}n^\beta \mathcal{D}\mathscr{C}_\nu^k   \mathcal{D}\mathscr{X}_\nu^b  \tilde{J} \delta(\tilde{\bm\chi})  
   \Delta_{\rm FP}^{\rm red}
   e^{-\tilde S_{\rm YM}[\bm n, \mathscr{C},\mathscr{X} ]}
\exp \left\{  ig_{{}_{\rm YM}} \sqrt{\frac{2(N-1)}{N}} [(j, N_{\Sigma_{C}}) + (k, \Xi_{\Sigma_{C}}) ] \right\}
 ,
\nonumber\\
 Z_{{\rm YM^\prime}}   
=&  \int \mathcal{D}n^\beta
   \mathcal{D}\mathscr{C}_\nu^k   
\mathcal{D}\mathscr{X}_\nu^b 
\tilde{J} 
\delta(\tilde{\bm\chi})  
   \Delta_{\rm FP}^{\rm red}
 e^{-  S_{\rm YM^\prime}[\bm n, \mathscr{C},\mathscr{X}] } . 
\end{align}

\noindent
Remark:
\begin{enumerate}
\item
For $SU(2)$, when we fix the color field $\mathbf{n}(x)=(0,0,1)$ or $\bm n(x)=\sigma_3/2$,  
 the reduction condition 
$D^\mu[\mathscr{V}]\mathscr{X}_\mu=0$
reduces to the conventional Maximally Abelian gauge \cite{KLSW87}.

\item
For $SU(3)$, this is not the case: This reduction does not reduce to the conventional Maximally Abelian gauge  for $SU(3)$, even if the color field is fixed to be uniform.
Therefore, the results to be obtained are nontrivial.
\end{enumerate}

\section{Conclusion and discussion}

We have combined a non-Abelian Stokes theorem for the Wilson loop operator \cite{Kondo08} and the new reformulations of the Yang-Mills theory \cite{KSM08} to study quark confinement from a viewpoint of the dual superconductor. 
The obtained results are summarized as follows. 
\\
\noindent
\begin{enumerate}
\item[1)] 
In order to define (chromo)magnetic monopoles in the $SU(N)$ Yang-Mills theory (without adjoint scalar fields),
 {we do not need to use the prescription called the \textbf{Abelian projection} ['t Hooft,1981]\cite{tHooft81}  which realizes magnetic monopoles as  \textbf{gauge-fixing defects}. 
 In fact, we can define  {gauge-invariant magnetic monopoles} which are inherent in the non-Abelian Wilson loop operator and we can extract them by using a \textbf{non-Abelian Stokes theorem} for the non-Abelian Wilson loop operator \cite{Kondo08}. 
 
\noindent
\item[2)]
For the $G=SU(2)$ gauge group, the resulting magnetic monopole coincide with one obtained from the CDG  decomposition for the Yang-Mills field which was proposed by [Cho (1980)] and [Duan \& Ge (1979)] independently.
} 
For the $G=SU(2)$ gauge group, such an Abelian magnetic monopole is described by the color field $\bm{n}(x)$ with the target space: $\bm{n}(x) \in SU(2)/U(1)=P^1(\mathbb{C})$ for quarks in any representation. 
However,  $G=SU(2)$ is an exceptional case.

\noindent
\item[3)]

For $SU(N)$ ($N \ge 3$), the resulting magnetic monopole depends on the representation of quarks defining the Wilson loop operator, which is related to the specific target space of the color field $\bm{n}(x)$.
For the $G=SU(3)$ gauge group, every  representation of $SU(3)$  is specified by the Dynkin index $[m,n]$ and the magnetic monopoles are exhausted by two cases:
\begin{itemize}
\item
For quarks in the representation $m=0 \ \text{or} \ n= 0$, $\tilde H=U(2)$ , e.g., the fundamental representation of $G=SU(3)$,  

a non-Abelian magnetic monopole described by $\bm{n}(x) \in SU(3)/U(2)=P^2(\mathbb{C})$

\item
For quarks in the representation $m \ne 0 \ \text{and} \ n \ne 0$, $\tilde H=H=U(1)\times U(1)$, e.g., the adjoint representation of $G=SU(3)$,   

two Abelian magnetic monopoles  described by $\bm{n}(x) \in SU(3)/[U(1)\times U(1)]=F_2$

\end{itemize}
Here $\tilde H$ is a subgroup of $G$ called the maximal stability group which is uniquely determined once the representation is specified. 
$\tilde H$ does not necessarily agree with the maximal torus group $H=U(1)^{N-1}$.


\noindent
\item[4)]
 We have constructed the new reformulations of the $SU(N)$ Yang-Mills theory using new field variables so that they give the optimal description of the  gauge-invariant  magnetic monopole defined through the $SU(N)$ Wilson loop operator. 
This is an extension of the work due to [Cho (1980)]\cite{Cho80c} and [Faddeev \& Niemi (1999)]\cite{FN99a} including their results as a special option where $N-1$ color fields $\bm{n}^{(j)}(x)$ ($j=1,...,N-1$) corresponding all the Cartan subgroup are introduced. 
The reformulation allows a number of options discriminated by the maximal stability group $\tilde{H}$ of the gauge group $G$.  
Our reformulations introduce  {only a single color field $\bm{n}(x)$ for any $N$}, which is enough for reformulating the quantum Yang-Mills theory to describe confinement of the  {fundamental quark}.  
For $G=SU(3)$, two options are possible: 
\begin{itemize}
\item
The maximal option with $\tilde{H}=H=U(1) \times U(1)$, the reformulation gives a manifestly gauge-independent extension of the conventional Abelian projection in the maximal Abelian gauge. 
This is just the case of Cho and Faddeev \& Niemi.

\item
The minimal option with $\tilde{H}=U(2)$ gives an optimized description of quark confinement through the Wilson loop operator in the fundamental representation.
[Kondo, Shinohara and Murakami, 2008]
The minimal option in our reformulation is new for $SU(N), N \ge 3$: 

\end{itemize}

\noindent
\item[5)]
Moreover, we have constructed the lattice versions\cite{KKMSSI06,IKKMSS07,SKKMSI07} \cite{KSSMKI08,SKS10} of the reformulations of the $SU(N)$ Yang-Mills theory and    performed numerical simulations on a lattice for $SU(2)$ and $SU(3)$.  [Talk by Akihiro Shibata, see the contribution to this conference.]

\noindent
5a) 
For $SU(2)$ and $SU(3)$, we have confirmed the infrared restricted field dominance and the magnetic monopole dominance for quark confinement: 
\begin{itemize}
\item
For $SU(2)$, we have confirmed the  {infrared dominance of the restricted variables} $\mathscr{V}$  (a gauge-independent version of the ``Abelian'' dominance) and the   {Abelian magnetic monopole dominance} for confinement of quark (in the string tension) in any representation. \cite{KKMSSI06,IKKMSS07}
\\
cf. [infrared Abelian dominance and Abelian magnetic monopole dominance in the MA gauge (Abelian projection)]

\item
For $SU(3)$, we have confirmed the  {restricted field dominance} $\mathscr{V}$   and  {the non-Abelian magnetic monopole dominance} for confinement of quark (in the string tension) in the fundamental representation. \cite{KSSK11}

\end{itemize}
For $SU(2)$ and $SU(3)$, we have presented the suppression of the remaining field $\mathscr{X}$ (exponential fall-off of the correlation function) in the low-energy or the long distance region. \cite{SKKMSI07}

\noindent
5b) 
For $SU(2)$ and $SU(3)$, we have given the numreical evidences for the  {dual Meissner effect caused by gauge-invariant magnetic monopoles} in the Yang-Mills theory:  simultaneous formation of the chromoelectric flux tube connecting a pair of quark and antiquark, and the magnetic current induced around the flux tube  \cite{SKKS13}. 

For $SU(2)$ and $SU(3)$, we have confirmed also the infrared restricted field dominance 
in the dual Meissner effect suggesting the magnetic monopole condensations. 

Moreover, we have determined the type of the dual superconductivity by measuring the penetration depth and the coherent length (assuming the relativistic Ginzburg-Landau model for fitting the data). \cite{KKS14,SKKS13}

\begin{itemize}
\item
For $SU(2)$, the type of the dual superconductivity is the border between type I and II or rather weakly type I \cite{KKS14}. This is consistent with the preceding works.

\item
For $SU(3)$, the type is strictly type I \cite{SKKS13}.  This is a new result which is consistent with the result of other groups.

\end{itemize}
These results support the  {non-Abelian dual superconductivity  picture for quark confinement in $SU(3)$ Yang-Mills theory}.

\end{enumerate}

\noindent
For applications of the reformulation to other topics, see the recent review \cite{KKSS14}.

\section*{Acknowledgements}
This work is financially supported in part by Grant-in-Aid for Scientific Research (C) 24540252 from Japan Society for the Promotion of Science (JSPS).
This work is in part supported by the Large Scale Simulation Program 
No.09-15 (FY2009), No.T11-15 (FY2011), No.12/13-20 (FY2012-2013) and No.13/14-23 (FY2013-2014) of High Energy Accelerator Research Organization (KEK).



\bibliographystyle{aipproc}   


\IfFileExists{\jobname.bbl}{}
 {\typeout{}
  \typeout{******************************************}
  \typeout{** Please run "bibtex \jobname" to optain}
  \typeout{** the bibliography and then re-run LaTeX}
  \typeout{** twice to fix the references!}
  \typeout{******************************************}
  \typeout{}
 }



\end{document}